\documentclass[lettersize, journal]{IEEEtran}
\usepackage{booktabs} % For formal tables
\usepackage[hidelinks]{hyperref}
\usepackage{lipsum}
\usepackage{graphicx}
\usepackage{siunitx, amsmath}

\usepackage{subfigure}
\usepackage{caption}
\usepackage{pifont}% http://ctan.org/pkg/pifont

\usepackage[normalem]{ulem}
\usepackage{multirow}
\usepackage{multicol}
\usepackage{comment}
\usepackage{color}
\usepackage{etoolbox}
\newbool{inccomment}
\setbool{inccomment}{true}
\newcommand{\hl}[1]{\ifbool{inccomment}{{\color{magenta}#1}}{}}
\newcommand{\fc}[1]{\ifbool{inccomment}{{\color{blue}#1}}{}}

\usepackage{bbm}

\usepackage{tabularx}
\usepackage{graphicx}
\usepackage{supertabular}
\usepackage{textcomp}
\usepackage{amssymb}

\usepackage{algorithm}
\usepackage{algpseudocode}
\algtext*{EndWhile}
\algtext*{EndIf}
\algtext*{EndFor}
\algtext*{EndProcedure}
\usepackage[super]{nth}
\usepackage[T1]{fontenc}

\usepackage{amsthm}
\usepackage{mathtools}
\usepackage{pdfpages}

\usepackage{url}
\usepackage{bbding}

\usepackage{dblfloatfix}

\newcommand\blfootnote[1]{%
  \begingroup
  \renewcommand\thefootnote{}\footnote{#1}%
  \addtocounter{footnote}{-1}%
  \endgroup
}

\usepackage{xcolor,colortbl}

\graphicspath{{./_fig/}}

\allowdisplaybreaks
\makeatletter
\def\BState{\State\hskip-\ALG@thistlm}
\makeatother
\begin{document}

\title{The Dark Side: Security Concerns in Machine Learning for EDA}

\author{Zhiyao Xie, Jingyu Pan, Chen-Chia Chang, Yiran Chen \textit{Fellow, IEEE} \vspace{-.3in}}

\maketitle

%=======================================%
%           Abstract          %
%=======================================%

\begin{abstract}

%The growing IC complexity has led to an increasingly compelling need for design efficiency improvement through new design automation methodologies. %A promising solution is to apply machine learning (ML) techniques in electronic design automation (EDA). However, while ML has demonstrated effectiveness in circuit design, the dark side about security problems, is seldomly discussed.
%In recent years, machine learning (ML) techniques have demonstrated great potential in electronic design automation (EDA). While most research efforts focus on .

The growing IC complexity has led to a compelling need for design efficiency improvement through new electronic design automation (EDA) methodologies. In recent years, many unprecedented efficient EDA methods have been enabled by machine learning (ML) techniques. While ML demonstrates its great potential in circuit design, however, the dark side about security problems, is seldomly discussed. This paper gives a comprehensive and impartial summary of all security concerns we have observed in ML for EDA. Many of them are hidden or neglected by practitioners in this field. In this paper, we first provide our taxonomy to define four major types of security concerns, then we analyze different application scenarios and special properties in ML for EDA. After that, we present our detailed analysis of each security concern with experiments. 

%In recent years, machine learning (ML) techniques have enabled many unprecedented solutions in EDA. 

%For each concern, we present our studies on them. 
%demonstrate our preliminary experiments and results on them. 

\end{abstract}

\blfootnote{Zhiyao Xie is with the Department of Electronic and Computer Engineering at the Hong Kong University of Science and Technology, Hong Kong SAR (email: eezhiyao@ust.hk).}
\blfootnote{Jingyu Pan, Chen-Chia Chang, and Yiran Chen are with the Department of Electrical and Computer Engineering at Duke University, Durham, NC 27708, USA (email: jingyu.pan@duke.edu; chenchia.chang@duke.edu; yiran.chen@duke.edu).}

\vspace{-.2in}

\section{Introduction}

%Electronic design automation (EDA) techniques have achieved remarkable progress over past decades. However, the current chip design flow is still largely restricted to individual point tools with limited interplay across different tools and design steps. Tools in early steps cannot well judge if their solutions may eventually lead to satisfactory designs, and the consequence of a poor solution cannot be found until very late. Such disjointedness in the design flow is traditionally mitigated by either simplified estimations with heuristics or iterative design, which often lead to over-conservative design or longer turn-around time, respectively.

Driven by the continuously growing complexity in integrated circuits (ICs), design companies are in increasingly greater demand for experienced manpower and stressed with unprecedented longer turnaround time. The nonrecurring engineering (NRE) cost associated with chip design also keeps skyrocketing accordingly~\cite{DesignCost}. Therefore, there is a compelling need for essential improvement on IC design efficiency through new methodologies and design automation techniques. To solve this, machine learning (ML) techniques are considered a highly promising direction.

In recent years, machine learning for EDA has become a trending topic~\cite{huang2021machine, rapp2021mlcad}. ML models are developed to improve the predictability in chip design flows, by providing early feedback on downstream design quality or accelerating the solution of EDA problems. These ML models learn from prior design solutions and typically perform orders-of-magnitude faster design quality evaluations or optimizations. 
We have witnessed ML solutions targeting various design objectives, covering all major design stages for both analog and digital designs~\cite{huang2021machine, rapp2021mlcad}. Some techniques are further adopted in commercial EDA tools~\cite{dsoai, cerebrus}. In both EDA academia and industry, ML for EDA has made an impressive impact. We have strong reasons to believe ML models will be more widely adopted in design automation in the future.

% make a more significant impact

 %These ML models learn from prior design solutions and typically perform orders-of-magnitude faster design quality evaluations or provide better design optimizations. 

Existing ML for EDA techniques seek various attractive properties, such as better design quality, shorter turn-around time, and a higher level of automation. A significant amount of research and engineering efforts have been invested in these targets. However, these properties are no longer desirable if fundamental \emph{security} requirements are not first satisfied. In this study, we use the term `security' in a broad sense to include all measures about causing and preventing unforeseen consequences.

\begin{figure}[!b]
\centering
    \vspace{-.05in}
    \includegraphics[width=\linewidth]{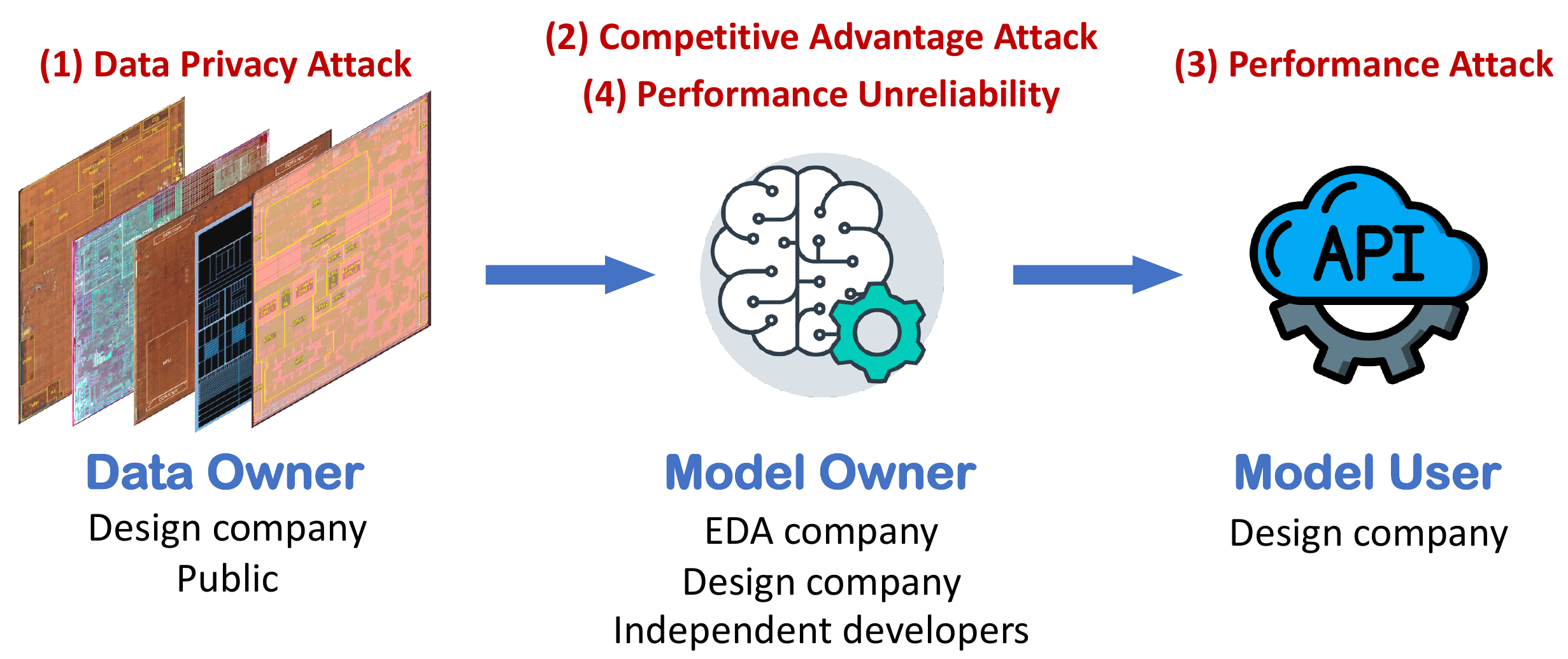}
  %  \vspace{-.25in}
\caption{Illustration of one typical ML for EDA flow.}
\label{overview}
\end{figure}

Actually, as ML is introduced in design automation, many unprecedented security concerns arise, but most practitioners are not fully aware of them. According to our study, the negligence of these potential security problems can lead to serious consequences for both model providers and users. 
Possible consequences include misleading results, design information leakage, model information leakage, etc. While a few previous works~\cite{liu2020adversarial, liu2020poisoning, yang2021attacking, liu2020bias} studied possible adversarial attacks on ML models targeting lithography problems, they only account for a very small portion of security challenges we observed in ML for EDA. In this paper, we try to give a more comprehensive and impartial study on all identified security challenges. We propose our taxonomy to define four major types of security concerns in ML for EDA: 

\begin{enumerate}
    \item \textbf{Attacks against data privacy}, e.g., attacks that try to infer private information about design data. 
    \item \textbf{Attacks against competitive advantage}, e.g., attacks that construct similar substitute models, which impair the competitive advantage of the original model. 
    \item \textbf{Attacks against ML performance}, e.g., adversarial or poisoning backdoor attacks that cause accuracy degradation on specific testing samples.
    \item \textbf{Inherent unreliability in ML performance}, e.g., unexpected accuracy degradation on new testing samples. 
\end{enumerate}

%These security concerns affect both the ML model and the design/IP as training data.

Figure~\ref{overview} illustrates a typical ML for EDA development and usage flow, and corresponding concerns. Malicious attacks are not the only source of security concerns in this paper. We also pay attention to unreliability problems, which cause unforeseen consequences in many scenarios and are especially serious in EDA and chip design. According to this taxonomy, all aforementioned previous studies~\cite{liu2020adversarial, liu2020poisoning, yang2021attacking, liu2020bias} can be categorized into the third type. In addition, a recent survey~\cite{rapp2021mlcad} on ML for EDA mentioned their concern about type-one attack on training data without a more detailed analysis.

In this paper, we present a comprehensive study with our preliminary experimental results on all identified security concerns. According to our observation, some of these security challenges are actually less practical, while others pose a high threat to data owners, model owners, or users. We will first present representative works, application scenarios, and special properties of ML for EDA in Section~\ref{sec:background}. After that, these four major security concerns are presented in Section~\ref{sec:1}, \ref{sec:2}, \ref{sec:3}, and \ref{sec:unreliability}, respectively. Finally, in Section~\ref{sec:end}, we discuss other potential concerns and impacts in the future, when ML for EDA becomes ubiquitous.

% Then we summarize all these concerns in Section~\ref{sec:summary}.

%In the remaining of this paper, we start with the ML for EDA background and their relations with security challenges in Section~\ref{sec:property}

%We can observe at least three types of security challenges on 1) \textbf{ML models}, 2) \textbf{design/IP}, 3) \textbf{EDA software}. Figure~\ref{overview} shows an overview of these security challenges. These challenges affect EDA vendors and design companies differently. 

%We focus the impact of these challenges on . 

%In the remaining of this paper, we will first study the background in ML for EDA and their relations with security challenges in Section~\ref{sec:property}. After that, our studies on the three different types of security challenges are introduced in detail in Section~\ref{sec:1}, Section~\ref{sec:2}, and Section~\ref{sec:3}, respectively.

\section{ML for EDA Background}
\label{sec:background}

\subsection{Existing ML Solutions in EDA}

We start with a brief inspection of representative ML solutions in EDA. Nowadays, ML-based research efforts can be observed at almost all major stages of a typical VLSI design flow. For high-level synthesis (HLS), models are proposed for fast quality of result (QoR) estimation~\cite{dai2018fast, makrani2019pyramid} or design space exploration~\cite{liu2013learning, liu2016efficient}. Many power models~\cite{zhou2019primal, zhang2020grannite} are also proposed in early design stages. Some power models~\cite{xie2021apollo, kim2019simmani} are further implemented for runtime circuit management. At logic synthesis, ML models are proposed for chip quality prediction~\cite{yu2018developing, neto2019lsoracle} and optimization~\cite{hosny2020drills, neto2021slap}. During physical design, more models perform predictions or optimizations on almost all important design metrics, including timing~\cite{barboza2019machine, kahng2015si, xie2021timing}, macro placement~\cite{mirhoseini2021graph, huang2019routability}, routability~\cite{xie2018routenet, yu2019painting, chen2020pros}, IR drop~\cite{xie2020powernet, ho2019incpird, zhou2020gridnet}, clock tree quality~\cite{lu2019gan}, interconnect~\cite{xie2021net}, crosstalk~\cite{liang2020routing}, 3D integration~\cite{lu2020tp}, etc. Also, ML models are developed for design verification~\cite{katz2011learning, fine2003coverage}, design for testability (DFT)~\cite{ma2019high}, and lithography problems~\cite{yang2018layout, yang2019gan}. Besides the methods applied at specific design stages, automatic design flow tuning is another well explored topic in ML for EDA~\cite{kwon2019learning, xie2020fist}.

ML-based methods are of course not only limited to digital designs. For analog design, similarly, various models have been developed for topology design~\cite{kunal2020general, li2016analog}, device sizing~\cite{wang2020gcn, hakhamaneshi2019bagnet}, pre-layout prediction~\cite{ren2020paragraph, kunal2020gana}, layout evaluation~\cite{liu2020towards, li2020customized}, layout generation~\cite{xu2019wellgan, zhu2019geniusroute}, and analog design testing~\cite{stratigopoulos2008error}. For a more complete survey on all existing research efforts, please refer to previous survey papers~\cite{huang2021machine, rapp2021mlcad} solely devoted to this topic.

Besides being a hot research topic in academia, ML-based estimators have also gained popularity in the EDA industry.
Recent versions of commercial tools already support the construction of ML models on delay~\cite{Innovus} or congestion predictions~\cite{ICC2}, providing improved PPA or faster convergence after invoking the ML models in their tools~\cite{Innovus, ICC2}. 
In addition, EDA vendors have provided ML models for design space exploration or design flow tuning, named DSO.ai~\cite{dsoai} and Cerebrus~\cite{cerebrus}.

Among these ML applications targeting digital or analog designs, almost all popular ML techniques have been applied.
Most methods in ML for EDA adopt supervised models, especially neural network techniques, while some others perform reinforcement learning. In this work, we also focus on the most popular supervised methods. 
Considering the popularity in both EDA academia and industry, we believe ML models will play a more important role in design automation in the future. Therefore, a deep understanding of all potential security concerns is essentially important.

%covering supervised, unsupervised, and reinforcement learning algorithms.

%In most recent works, a strong trend towards neural network (NN)-based algorithms, especially deep learning techniques, can be observed~\cite{rapp2021mlcad}. 

% In summary, ML for EDA has made an impressive impact in both EDA academia and industry.

%\subsection{ML-Related Security Problems}
%No systematic studies on security problems have been . 
%However, chip design. Most practitioners are not ready for 
%as ML methods gain increasing popularity in design automation, 

%\vspace{-.05in}
\subsection{Application Scenarios}

To better analyze all security concerns in ML for EDA, we should first fully understand the practical applications scenarios of these ML-based techniques. However, as an emerging type of chip design technique, new explorations in ML for EDA solutions are still ongoing while the pace of commercialization in the industry lags behind. Thus, besides observing existing solutions, we have to anticipate possible application scenarios in near future.

Currently, many existing research efforts in ML for EDA merely target the demonstration of their correctness and effectiveness. A small portion of works have been verified and applied in private in-house design flows in design companies. In addition, some ML models are deployed in EDA tools by EDA vendors. 
They correspond to two major types of application scenarios. 
1) Same model providers and users. For ML models developed and deployed for in-house flows internally, the model provider and user are from the same company and work rather closely. 2) Separate model providers and users. As Figure~\ref{overview} shows, there may be separate model providers from EDA vendors or independent developers and model users from design companies. 
In the future, we tend to believe it is more likely for more ML model providers and users to be separated, like the separation of IC design, EDA, and fabrication in the semiconductor industry history.

%Therefore, there exist two major types of application scenarios nowadays. 
%Just like the separation of IC design, EDA, and fabrication in semiconductor industry history, we tend to believe it is more likely for more ML model providers and users to be separated in the future. 

Despite these observations, the anticipation of future application scenarios is not straightforward. Compared with traditional EDA software, ML for EDA methods adopt a different and more complex flow, which consists of multiple stages, including model architecture design, data and label collection, model training, model inference/prediction, and utilization of prediction results. These tasks could be divided differently between model providers and clients. Different partitions of tasks lead to different scenarios.

%\textbf{Provide black-box trained ML model.} 
%\textbf{Provide black-box untrained ML model.} 
%\textbf{Provide white-box trained ML model.} 

\begin{table}[!b]
  \centering
 % \vspace{.06in}
  \caption{Possible Application Scenarios of ML for EDA}
  \label{tbl:scenario}
  \resizebox{\linewidth}{!}{ \renewcommand{\arraystretch}{1.1} 
  \begin{tabular}{l | c c c}
 	\hline
    Scenario & Black-Box &  Trained &  Separated Provider \& User \\        
	\hline
	S1  &  \Checkmark    &  \Checkmark   &  \Checkmark   \\ 
	S2  &  \XSolidBrush  &  \Checkmark   &  \Checkmark   \\ 
	S3  &  \Checkmark    &  \XSolidBrush &  \Checkmark   \\
	S4  &  \XSolidBrush  &  \Checkmark   &  \XSolidBrush \\
	\hline
  \end{tabular}
  }
%      \vspace{-.25in}
\end{table}

Table~\ref{tbl:scenario} presents four possible application scenarios or business models of ML for EDA based on our anticipation. In the first scenario S1, a separate ML model provider provides their well-trained model as a black-box to users, possibly through cloud services. This is very similar to the popular ML-as-a-service (MLaaS) business model in many general ML tasks, like the cloud services offered by Amazon, Google, Microsoft, BigML, etc~\cite{tramer2016stealing}. Such cloud services allow model providers to charge users for queries. These ML models are of high commercial values. In this case, models will be vulnerable to many possible malicious attacks. 

%A possible scenario or business model of ML for EDA is the S1, 

In addition, there could be a special case, S2, where ML models are actually white-box to users or potential attackers. There are a few possible reasons causing the model to be white-box. For example, researchers, individual developers, and even companies may hope to directly open-source their trained model for free. Also, models targeting black-box in S1 may be hacked, especially if they are deployed locally instead of through cloud platforms without enough security measurements. In this scenario, the ML model itself is already available to potential attackers, while new security concerns about the training data arise.

Another possible scenario, S3, is to leave more tasks to users. The model providers only design their ML methodology without performing the training. The method is provided as black-box, with information like feature, architecture, and optimization procedure not explicitly disclosed. Then users can train and use their own customized ML models as black-box with their own labeled data. Rather than being provided as stand-alone services in S1, it is more likely for such methodologies to be integrated and released together with existing EDA tools. This business model can already be observed in some existing EDA tools~\cite{Innovus} from vendors.

Finally, ML model providers and users may not be separated. Users in design companies can design and train their own models for specific problems in their in-house design flow. This is scenario S4 in Table~\ref{tbl:scenario}. In this case, this rather private flow will be much less vulnerable to malicious attacks. But it will still be affected by the inherent unreliability of ML models, which will be covered in detail in Section~\ref{sec:unreliability}.

%\vspace{-.1in}
\subsection{Overview of Special Properties}
\label{subsec:property}
%\subsection{Overview of Security Challenges}
%Although ML for EDA algorithms share many similarities with ML applications in other fields, there are also some special properties of these methods. These properties lead to several serious securities challenges, which are not yet systematically studied.

Before giving a detailed analysis of security challenges, we briefly inspect some special properties of ML for EDA solutions. Although many ML for EDA solutions have been developed based on black-box use of existing ideas from the ML community, we still observe some remarkable properties different from general ML tasks. 

%related to security studies. 

%of most ML for EDA methods. %and they contribute to security vulnerabilities. %Some of these properties are actually not common in other ML applications. 
%These properties, as shown below, contributed to several serious securities challenges. 

%\textbf{Unprecedented data heterogeneity and complexity.} A circuit contains orders-of-magnitude more information than an image and the difference between circuits due to micro-architecture, functionality, and technology node is significantly larger than the difference between general images. 

%Due to such heterogeneity in circuit data, the model accuracy can significantly degrade when tested on a largely different new circuit.

\textbf{Unprecedented data heterogeneity.} Huge heterogeneity can exist between data samples, resulting from the large difference among circuit designs due to functionality, micro-architecture, and technology node. 
For example, assuming we already restrict the training and testing data of an ML model to be only from Arm processors, we still cannot expect the model trained on old designs like Cortex-M0 with 40nm technology node to perform very well on latest designs like Neoverse N2 with 5nm technology. 
This level of training and testing data heterogeneity is uncommon in benchmarks for general ML applications like computer vision.

\textbf{High complexity in data and pattern to learn.} A circuit contains orders-of-magnitude more information than an ordinary image. For prediction tasks, models are learning behaviors of highly complex EDA engines. For optimization tasks like macro placement, models are exploring a huge solution space~\cite{mirhoseini2021graph}, significantly larger than the Go game solved by AlphaGo~\cite{silver2016mastering}. These complexities increase the difficulty in studying security problems in ML for EDA.

%These complexities 

%models trained on designs like Cortex-M0 with 90nm technology may not perform well on latest processors, like Neoverse N2 with 5nm technology. 

\textbf{More confidential design in higher demand.} The construction of ML models in EDA relies on training data generated from circuit designs, which are highly confidential to design companies. Due to the aforementioned data heterogeneity, for ML models targeting most cutting-edge circuit designs, similarly latest cutting-edge circuits are typically desired as training data for model construction. This tends to put these advanced highly confidential circuit designs at a higher risk of information leakage.

\textbf{Potentially decentralized training data.} Many ML for EDA developers have very limited access to the latest design data owned by design companies. Therefore, training with decentralized private circuit data is explored in recent works~\cite{pan2022towards}. They propose to perform collaborative training on decentralized data with techniques like federated learning~\cite{mcmahan2017communication}. Such a scenario can lead to many additional risks.

%Those most cutting-edge advanced design data is desired for . 

%data leakage risk and unreliability of ML model performance. 
%Although one can train ML models within a design company, the data of a single company might still be inadequate or biased, especially for small companies. 

\textbf{Model performing binary classification or regression.} Most security studies in general ML tasks target common multi-class classifiers. For example, there are 1000 classes in Image-Net benchmark for convolutional neural network (CNN) models and 3 classes in COLLAB benchmark~\cite{yanardag2015deep} for graph neural network (GNN) models. In comparison, most predictive models in EDA perform binary classification or regression, while optimization models adopt reinforcement learning. This difference makes many attack and defense methods targeting multi-class classifiers no longer applicable.

%\fi

%\textbf{Commercialized EDA tools for label generation.} Most supervised ML applications rely on human experts to annotate training data to generate labels. These models can replace human in many scenarios. In comparison, commercial EDA tools are commonly employed to generate labels in ML for EDA. As a result, ML models can potentially replace many functions of these commercial tools, introducing unprecedented competitions in the market. 

%In summary, these properties contribute to securities challenges in different ways. First, security risks in design and IP are naturally caused by the high confidentiality and the decentralized distribution of training data. Third, the security risks of ML models in EDA are contributed by both confidentiality of training data and the data heterogeneity. 

%In summary, the high confidentiality in training data naturally leads to security risks in design and IP. The decentralized distribution of training data causes security risks in both design and ML models.   

%We observe various security vulnerabilities in ML for EDA, in both model development and deployment scenarios. In summary, ML methods will significantly change the . 

%\vspace{-.05in}
\subsection{Overall Experiment Setup}

In this paper, we try to cover all security concerns we observe in ML for EDA and provide our studies based on a few representative datasets. We perform our experiments mainly on the routability problem, a well-studied topic in ML for EDA. 
 %Extensive experiments on more ML for EDA topics are devoted to our future works. 

Previous routability estimators use either routing congestions~\cite{chen2020pros, yu2019painting} or DRC (design rule checking)~\cite{xie2018routenet, liang2020drc} as the metric of routability.
They detect congestion locations or DRC hotspots locations.
Given a set of placement solutions with extracted input feature maps $X$, routability estimators generate a neural network model $f$ to detect the locations of DRC hotspots or congestions $y$:
%\vspace{-.02in}
$$
f: X\in \mathbb{R}^{d\times h\times c} \rightarrow y
\in \{ 0, 1 \}^{d \times h} \vspace{-.02in}
$$
where $d$ and $h$ are the width and height of the layout, and $c$ indicates the number of input features/channels. 

In routability prediction tasks, congestion detection is simpler than DRC violation detection in practice. Therefore, congestion detection models generally achieve higher accuracy.

Most experiments in this work are based on a comprehensive dataset using 74 designs with largely varying sizes from multiple benchmarks.
There are 29 designs from ISCAS’89~\cite{brglez1989combinational}, 13 designs from ITC'99~\cite{corno2000rt}, 19 other designs from Faraday and OpenCores in the IWLS'05~\cite{albrecht2005iwls}, 13 designs from ISPD'15~\cite{bustany2015ispd}.
For each design, multiple placement solutions are generated with different logic synthesis and physical design settings.
Altogether 7,131 placement solutions are generated from these 74 designs.
We apply Design Compiler\textsuperscript{\textregistered} for logic synthesis and Innovus\textsuperscript{\textregistered}~\cite{Innovus} for physical design, with the NanGate 45nm technology library~\cite{URL:NanGate}.

Besides routability tasks, we also conduct experiments on lithography hotspot detection, another representative topic in ML for EDA, to study relevant security concerns on adversarial attacks. The lithography hotspot detectors are also CNN-based. The experiment is based on a lithography dataset from the previous work~\cite{yang2021attacking}, with four groups of 400 hotspot clips for adversarial sample generation and 34356 layout clips for model training. 

%Detailed setup can be found in many previous works~\cite{yang2021attacking}.

\section{Attacks Against Design Privacy}
\label{sec:1}

%We observe three major categories of challenges in ML for EDA. 

\subsection{Design Privacy Overview}

Training data is the foundation of ML for EDA and it directly determines the quality of ML models. Such data includes both input features and ground-truth labels. For a circuit design/IP used for data generation, input features are different representations of the design, and labels are corresponding circuit qualities including power, performance, etc. A circuit is significantly more complex than an ordinary image, thus provides rich information for model development. Such information can be highly confidential for design companies.

Previous studies~\cite{carlini2019secret, fredrikson2015model} have demonstrated that given an ML model, it is possible for attackers to reconstruct or recover sensitive feature information in the model training data. The process of malicious recovering input features is commonly referred to as \emph{model inversion} or \emph{reconstruction attack}~\cite{fredrikson2015model}. In ML for EDA, such attacks may cause serious security challenges on circuit designs/IPs used in training. Even compared with other ML applications involving private data, like medical image processing or language models on smartphones, attacks targeting ML for EDA models are more threatening, since attackers do not require high-quality recovery of training data. A very small part of information about the circuit design may already benefit the attacker. For example, attackers may only target basic information like dynamic scaling granularity, target manufacturing process, flat/horizontal implementation methodologies, etc. Based on the small piece of reconstructed features, it is possible for attackers with sufficient background to infer valuable information about the research or development direction of their target company.

To make things worse, as mentioned in Subsection~\ref{subsec:property}, in ML for EDA, due to data heterogeneity, more confidential design is in higher demand as high-quality training data. This property tends to put those most advanced and confidential circuit designs at a high risk of information leakage. This concern on design privacy is recognized as an open challenge by the recent ML for EDA survey~\cite{rapp2021mlcad}. 

% Although there are few studies on it, 

%To make things worse, huge difference exists between circuits due to micro-architecture, functionality, and technology node. Therefore, ML in EDA has a high requirement on the `quality' of training data. To construct high-performance ML models on cutting-edge industrial designs, it is commonly believed that the model should be trained with similar or even the same design, and the similar technology node. For example, assuming we develop ML models for Arm Ltd, we cannot expect a model trained on some old designs like Cortex-M0 with 90nm technology node to perform very well on latest designs, like Neoverse N2 with 5nm technology. More detailed discussions are provided in the Section~\ref{sec:unreliability}. This property tends to put those most advanced and confidential circuit designs in a high risk of information leakage.  

% Thus high-quality training data often comes from the most advanced and confidential designs for semiconductor companies.

% Such `quality' refers to the level of design complexity and target technology node. 

\subsection{Attack Method on Design Privacy} 

We provide a demonstration of the malicious reconstruction of training data in ML for EDA. It applies to most complex ML models like deep neural networks. However, it turns out that such an attack has very high requirements on information available to attackers.

The fundamental attacking mechanism is straightforward. Given a differentiable ML model $F$ with trained weights $w$, denote the input features and the label of a training sample as $X$ and $y$, respectively. Then the trained model $F$'s prediction of this training sample $(X, y)$ can be denoted as $p$, which equals $F (X | w)$. 

%and $N$ training samples with features $X_i$ and label $y_i$, where $i \in [1, N]$. Then the trained model's prediction of the $i^{\text{th}}$ training sample can be denoted as $p_i$, which equals $F (X_i | w)$. 

%When the attacker can access the model as white-box, as indicated by scenario S2 in Table~\ref{tbl:scenario}, he has full knowledge of the weights $w$. However, the information about the ML model itself is not enough. We apply a very strong assumption to study the most threatening case of such an attack. If an attacker targets the $i^{\text{th}}$ training sample, we assume he/she can generate or hack a close estimation of the model prediction $p'_i \approx p_i = F(X_i | w)$ of this sample. This assumption is also made in representative reconstruction attack works~\cite{fredrikson2015model} on image models. Based on this, attackers can try to reconstruct similar input features of sample $i$, denoted as $X_{ai}$, targeting $X_{ai} \approx X_i$. This $X_{ai}$ can be referred to as \emph{reconstructed input}. The attacking process starts with an initial generation of the $X_{ai}$ with random signals. After that, gradient descent with respect to $X_{ai}$ is performed iteratively, as shown below, until it reaches convergence. 

When the attacker can access the model as white-box, as indicated by scenario S2 in Table~\ref{tbl:scenario}, he has full knowledge of the weights $w$. However, the information about the ML model itself is not enough. We apply a very strong assumption to study the most threatening case of such an attack. If an attacker targets the training sample, we assume he/she can generate or hack a close estimation of the model prediction $p' \approx p = F(X | w)$ of this sample. This assumption is also made in representative reconstruction attack works~\cite{fredrikson2015model} on facial image models. Based on this, attackers can try to reconstruct similar input features of the sample, denoted as $X_r$, targeting $X_r \approx X$. This $X_r$ can be referred to as \emph{reconstructed input}. The attacking process starts with an initial generation of the $X_r$ with random signals. After that, gradient descent with respect to $X_r$ is performed iteratively, as shown below, until it reaches convergence. 

\begin{align}
    %\text{In each iteration, \ }   X_{ai} \ \  -&= \ \nabla_{X_{ai}} Loss (F(X_{ai} | w), \ p'_i) \nonumber \\
    % & = \ \nabla_{X_{ai}}  || F(X_{ai} | w) - p'_i ||_2
    \text{In each iteration, \ }   X_r \ \  -&= \ \nabla_{X_r} Loss (F(X_r | w), \ p') \nonumber \\
     & = \ \nabla_{X_r}  || F(X_r | w) - p' ||_2
     \label{eq:input_opt}
\end{align}

Different from the model training process, where gradient descent is performed with respect to model weights $w$, in this attack, gradient descent is performed with respect to the reconstructed input $X_r$. This operation minimizes the difference between the prediction $F(X_r | w)$ based on attacker-reconstructed input $X_r$ and the actual prediction $p$ based on $X$. By performing this, ideally $X_r$ is optimized to approximate the original training data sample $X$.

However, in practice this simple loss function $|| F(X_r) - p' ||_2$ does not work well. Simply minimizing the difference between original and new model output may not optimize the reconstructed input $X_r$ towards the original feature $X$. This is also verified in our own experiment. Instead, extra loss function terms have to be introduced to steer the optimization direction and enforce the similarities between $X_r$ and original feature $X$~\cite{yin2020dreaming}.

To improve the attack quality, we provide additional guidance to make the $X_r$ follow existing feature statistics, which are stored in widely-used batch normalization (BN) layers of deep neural networks. This is inspired by the work of~\cite{yin2020dreaming} in computer vision. The BN layer~\cite{ioffe2015batch} normalizes the feature maps during training and implicitly captures the channel-wise running/moving means $\mu_{BN}$ and variances $\sigma^2_{BN}$. 
Therefore, we can steer the mean $\mu$ and variance $\sigma^2$ of input batches with reconstructed input $X_{a}$ towards the running values stored in all BN layers. 
We define regularization terms for the $l^{th}$ BN layer with $\mu_{BN\_l}$ and $\sigma^2_{BN\_l}$, as shown below.

%We can assume that feature statistics follow the Gaussian distribution across batches with mean $\mu$ and variance $\sigma^2$. 

\vspace{-.1in}
\begin{equation*}
    R ^l (X_r) = ||\mu_l(X_r) - \mu_{BN\_l}||_2  + ||\sigma_l^2 (X_r) - \sigma^2_{BN\_l} ||_2 
\end{equation*}

\noindent where $\mu_l(X_r)$ and $\sigma_l^2 (X_r)$ are the mean and variance of the batch with reconstructed input $X_r$ at the $l^{th}$ BN layer. Then these penalty terms corresponding to all BN layers are added to the loss function, with a controllable weight $\alpha$. 

\vspace{-.1in}
\begin{equation}
    Loss(F(X_r|w), p') = ||F(X_r|w) - p'||_2 + \alpha\sum_l R^l(X_r)  \\
    \label{eq:input_loss}
\end{equation}

%\vspace{-.1in}

%define the loss function to also minimize the ... in batch normalization layers.   

\begin{figure}[!t]
\centering
    \subfigure[]{\includegraphics[height=0.44\textwidth]{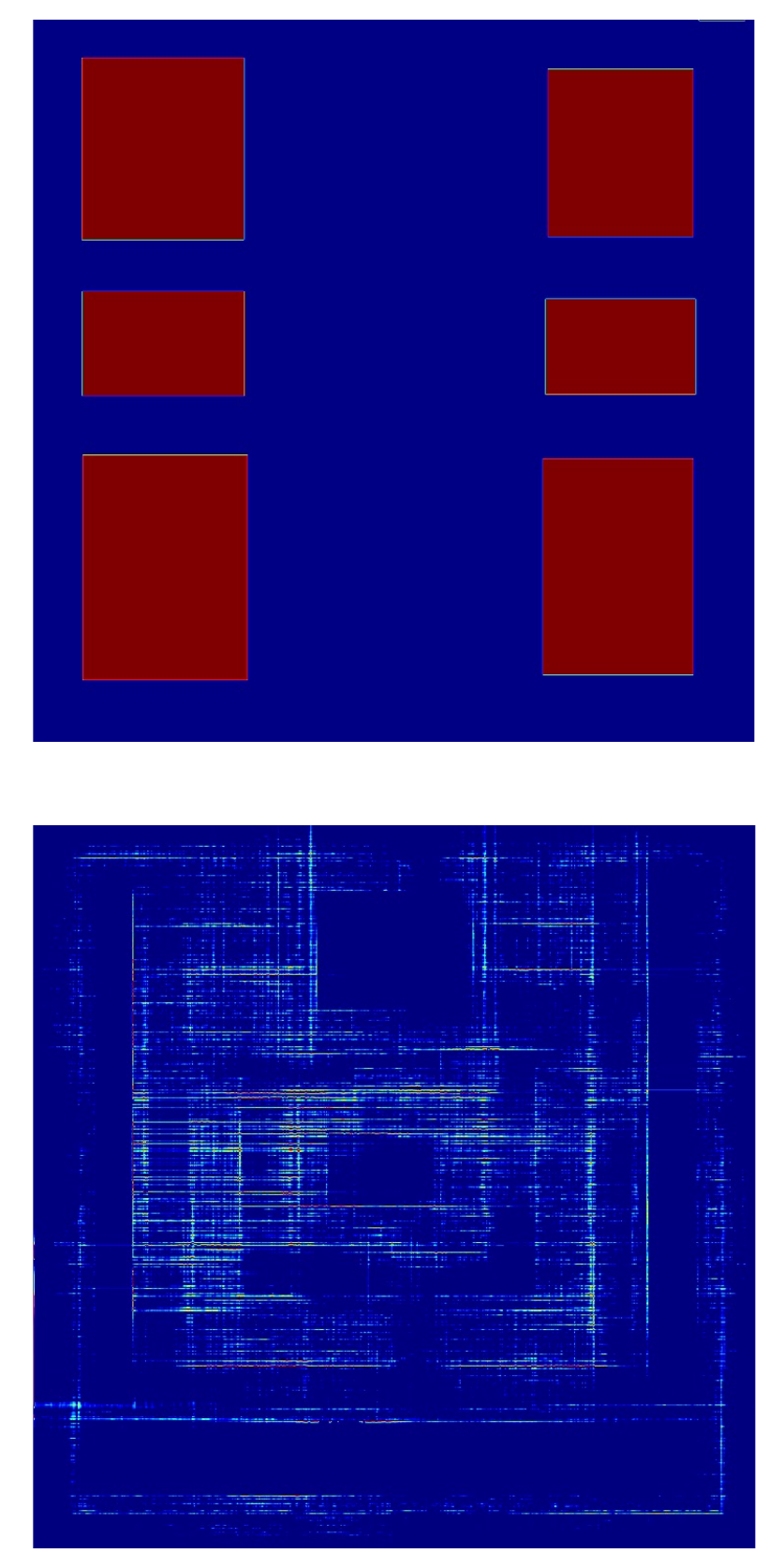}
    \label{inversion_l}}
    \hspace{.02in}
    \subfigure[]{\includegraphics[height=0.44\textwidth]{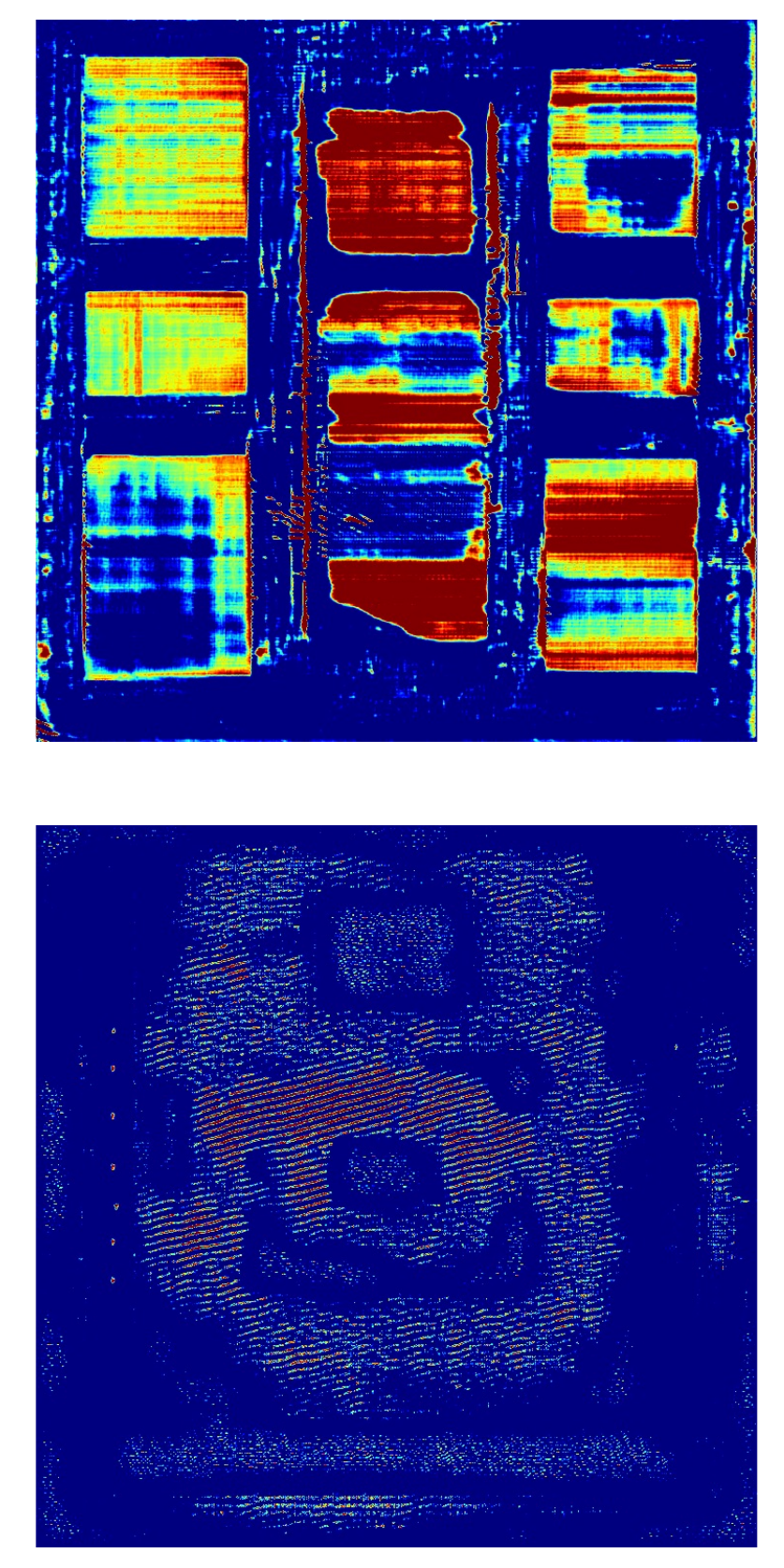}
    \label{inversion_p}}
    \vspace{-.05in}
\caption{Malicious model inversion attack. (a) The feature $X$ in training data (left column). (b) The reconstructed input $X_{a}$ (right column).  Target features on: the distribution of macros ($1^{st}$ row) and net bounding boxes ($2^{nd}$ row). }
\label{inversion}
    \vspace{-.1in}
\end{figure}  

\noindent In this way, the extra regularization steers the optimization of $X_r$ towards the recovery of original training features $X$.

\subsection{Experiment on Design Privacy Attack}

%Our experiment demonstrates that attackers can learn about the layout with their recovered feature $X_{ai}$. 

Experimental results on model inversion attack with loss in Equation~\ref{eq:input_loss} are shown in Figure~\ref{inversion}. We present our inversion results on two routability prediction features: macro positions and density of all net bounding boxes. 
The original features are shown in Figure~\ref{inversion_l} and the features reconstructed by attackers are in Figure~\ref{inversion_p}. A certain degree of similarities can indeed be observed, especially in large-scale patterns. For the macro locations, the sizes and locations all six macros in Figure~\ref{inversion_l} are reconstructed by $X_a$ in Figure~\ref{inversion_p}. For net bounding boxes, similarly, the regions with high net density are reconstructed. However, obvious differences still exist in both large-scale and small-scale patterns. For macros in Figure~\ref{inversion_p}, three false-positive macros are generated in the middle. Similarly, there are also false-positive net bounding box densities reconstructed in originally empty regions. 

More importantly, we emphasize that such attack is already based on a few very strong assumptions: (1) Attacker has white-box access to the ML model; (2) Attacker has an approximation of the prediction value $p'$. While the first condition may be achieved by hacking in scenario S2 or building very similar surrogate models, the generation of $p'$ is very difficult in practice. Despite these strong assumptions and the carefully designed attack algorithm in Equation~\ref{eq:input_loss}, we still get limited performance on design privacy attack, as reflected in Figure~\ref{inversion}. Therefore, we conclude that based on existing techniques and our current exploration, the overall difficulty to conduct a model inversion attack on design privacy in ML for EDA is actually high. 

% on white-box model 
%It shows the comparison between t

\section{Attacks Against ML Model Competitiveness}
\label{sec:2}

\subsection{ML Models Competitiveness Overview}

As indicated by scenario S1 in Table~\ref{tbl:scenario}, trained ML models can be provided on the cloud as a service in ML for EDA. Such MLaaS typically charges clients based on their queries. For service providers, it takes extensive efforts to construct these high-quality models, with steps including data collection, label generation, ML model design, ML model training and validation, etc. To provide even better service, they may have to construct multiple ML models for different types of design and technologies, taking extra engineering efforts. In summary, these trained ML models are important business assets and are costly to develop. 

%provide their core technical competence. 

However, it is possible for attackers to `steal' these models. Here the `steal' broadly refers to all activities where attackers develop their own substitute ML models with very similar functionality, utilizing the existing model in MLaaS. In other words, based on an existing black-box model $F$, attackers can train their own model, named \emph{attack model} $F_a$, with much lower cost. This malicious attack is referred to as \emph{model extraction}. Although this attack does not affect the function of the original MLaaS, the attack model $F_a$ poses an obvious threat to the competitive advantage and business value of the original model $F$.

%Besides attacks targeting trained models in scenario S1, . 

In addition, aforementioned scenario S1 in Table~\ref{tbl:scenario} is not the only vulnerable business model. In scenario S3, where only ML model architecture is provided as black-box without performing the training, malicious attacks are also possible. Attackers may infer the model architecture, in order to save their own research cost. In general ML applications, this has been achieved by building an extra ML model to map from the concatenation of query outputs to the model architecture attributes~\cite{oh2018towards}. It can be further improved by crafting own training data that maximizes information leakage. However, this attack on model architecture has only been verified on very simple models with less than 5 convolutional layers~\cite{oh2018towards}.  

% As indicated in previous works~\cite{chang2021auto} on routability prediction, model architecture can significantly affect the performance. 

%such that saving the development cost. 
%and build their own ML model.   

\subsection{Attack Method on Model Competitiveness}
\label{subsec:competitiveness}

For attackers who hope to build their own attack model $F_a$ in scenario S1, they can actually greatly benefit from existing trained ML models. The most fundamental yet effective attack methodology is to generate pseudo labels by querying the MLaaS-provided model $F$ with attackers' own unlabeled data. In practice, label generation is one of the most costly steps during model development in ML for EDA. First, it can take a large computation cost and long runtime to finish a design flow and get accurate simulation results, which are the labels. For example, assuming we work on a design with more than one million gates, it easily takes more than one day to finish synthesis and physical design to generate one complete layout. If developers plan to generate 1,000 labeled samples on designs at this level of complexity, it will take dozens of machines running for months. Second, this label generation process requires licenses of commercial tools. Third, it requires great engineering expertise and efforts to generate reasonable and realistic training labels. In summary, label generation requires extensive computation resources, commercial EDA tool licenses, engineer efforts, time, etc. 

If potential competitors/attackers can skip the label generation process to build their own dataset, the model construction will be much easier. We refer the provided MLaaS black-box model as the \emph{victim model} $F$ with trained weights $w$ and the ML model developed by attackers as the \emph{attack model} $F_a$ with weights $w_a$. Given unlabeled input data $X_u$, the attacker can query the victim model $F(X_u | w)$ and use it as the pseudo label to train the attack model $F_a$. So the attack method is a very simple gradient descent optimization, as shown below.

% And this is possible by utilizing provided MLaaS models. 

\vspace{-.1in}
\begin{equation*}
    \text{In each iteration,\ } w_a \ \  -= \ \nabla_{w_a} Loss (F_a(X_u | w_a), \ F(X_u | w)) 
\end{equation*}

This is the most fundamental while effective attack targeting scenario S1. Based on this, attackers may further reduce the number of queries, in order to save the cost. For example, they can choose to select and only query the most representative unlabeled samples, based on ideas from active learning or semi-supervised learning.

%by selecting the most re. 

\begin{figure}[!t]
\centering
    \includegraphics[width=\linewidth]{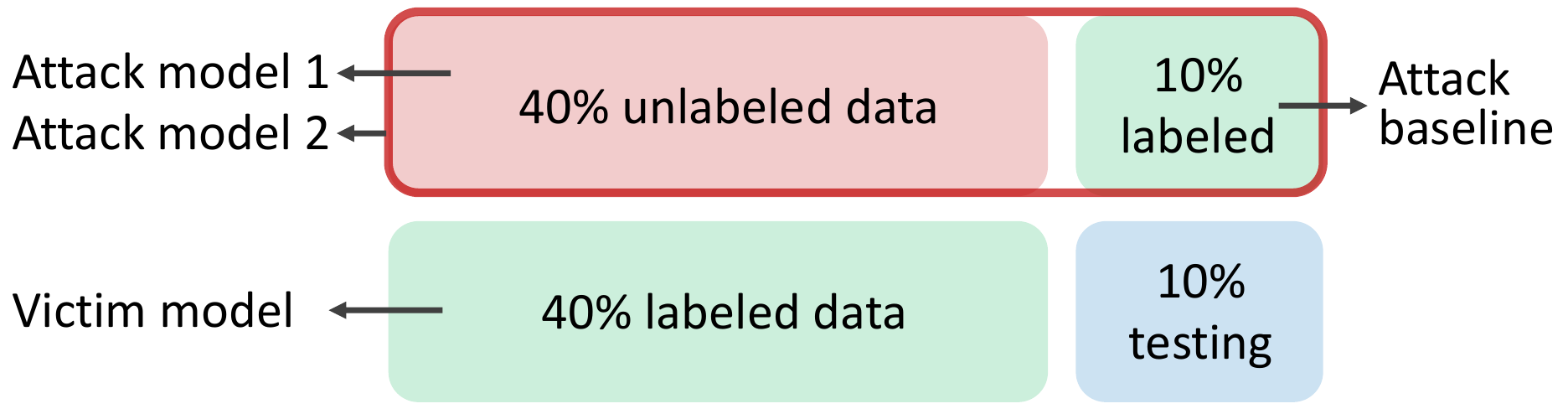}
    \vspace{-.1in}
\caption{The partitioning of dataset to study model competitiveness attack. Performance of these models are in Table~\ref{tbl:model_compete}.}
    \vspace{-.1in}
\label{partitioning}
\end{figure}

\subsection{Experiment on Model Competitiveness Attack}
\label{subsec:competitiveness_exp}

We demonstrate the effectiveness of our proposed fundamental model extraction attack in the routability experiment on constructing congestion models. Following the aforementioned scenario, we divide all of our existing data into four partitions without any overlap: 1) 40\% of labeled data used to train the original victim model. 2) 10\% labeled data used for testing model accuracy. 3) 40\% of unlabeled data prepared by attackers. 4) 10\% labeled data prepared by attackers, in order to build a baseline. Figure~\ref{partitioning} illustrates the data partitioning in this setup. Notice that the 40\% unlabeled data from attackers are different from the 40\% labeled training data of the victim model. This is very close to a realistic scenario, where attackers use different data from model developers.  

Based on the data partition, an attack baseline is first trained with 10\% labeled data. Then an attack model 1 is trained on 40\% unlabeled data with pseudo labels from victim model $F$. No actual label is provided by the attacker for this model. An attack model 2 is trained on both 40\% unlabeled data and the 10\% labeled data.

Table~\ref{tbl:model_compete} shows performance comparisons between the original victim model $F$, attack model baseline, and two attack models $F_a$. For attack model 1, without any labeled data, it achieves an accuracy of AUC=0.796, which is close to the victim model. For attack model 2, with a small portion (10\%) of extra labeled data, it achieves even higher accuracy (AUC=0.811) than the victim model. These results demonstrate the effectiveness of model extraction attack with such a simple pseudo-labeling method.

According to the result in Table~\ref{tbl:model_compete}, attackers can train even more accurate models with a very small portion of labeled data by querying the victim model. This attack proves to be efficient and profitable. It poses a threat to the competitiveness and business value of provided models in ML for EDA.

\begin{table}[!t]
  \centering
    \vspace{.05in}
%  \resizebox{\linewidth}{!}{%
\resizebox{0.95\linewidth}{!}{
  \begin{tabular}{| c | c | c |}
  \hline
\multirow{2}{*}{Model}   &  \multirow{2}{*}{Training data}  & Accuracy     \\
                         &                                 &  (AUC)       \\
  \hline
  \hline
  \multirow{2}{*}{MLaaS-provided victim}   & \multirow{2}{*}{40\% labeled data}   &   \multirow{2}{*}{0.806}    \\  
   &   &  \\
 % MLaaS-provided victim  &  40\% labeled data  &  0.806 \\
  \hline
  \multirow{2}{*}{ Attack baseline}  & \multirow{2}{*}{10\% labeled data }  &  \cellcolor{red!25}   \\
    &   &  \multirow{-2}{*}{\cellcolor{red!25}  0.765}  \\
  \hline
  \hline
  \multirow{2}{*}{Attack model 1}    & \multirow{2}{*}{40\% unlabeled data} & \cellcolor{green!25} \\
                    &   &   \multirow{-2}{*}{\cellcolor{green!25} \textbf{0.796} }  \\
  \hline
  \multirow{2}{*}{Attack model 2}   & 40\% unlabeled data  &  \cellcolor{green!25}   \\
                    & + 10\% labeled data   &   \cellcolor{green!25} \multirow{-2}{*}{ \textbf{0.811}}   \\
  \hline
  \end{tabular}
  }
  \vspace{.1in}
 \caption{Attack on model competitiveness. The MLaaS-provided (victim) model and attackers use different data.}
   \vspace{-.1in}
  \label{tbl:model_compete}
\end{table}

%  by constructing their own models with queries

% by extracting the ML models by sending queries. 
%The malicious attacks on ML for EDA may target either . 
%\subsubsection{Malicious Attacks}

%Most latest design data are owned by design companies and are highly confidential.  

\section{Attacks Against ML Performance}
\label{sec:3}

%\subsection{Adversarial and Poisoning Attack}
\subsection{ML Performance Attack Overview}

Besides aforementioned attacks targeting data privacy or model competitiveness, another main type of malicious attacks may happen in ML for EDA targets affecting the performance of existing ML models. Compared with the previous two types of attacks, which are less explored by ML for EDA community, some prior works~\cite{liu2020adversarial, liu2020poisoning, yang2021attacking} studied the attack on the performance of CNN-based lithography hotspot detectors.

There exist multiple types of malicious attacks on the performance of ML models. A well-studied type is adversarial attack, where attackers modify the model input by very small but deliberate alterations, named adversarial perturbation. In this way, attackers introduce their desired misleading ML inference result, without being noticed by potential victims. Such adversarial perturbation makes use of the inherent susceptibility of deep neural networks. However, in practice, it may not be feasible for outside attackers to easily modify the input in an ML-integrated circuit design flow.

The work of~\cite{liu2020adversarial} presents a realistic scenario of adversarial attacks on ML models targeting lithography hotspot detection. Currently, using a CNN-based hotspot detector, the designer can quickly ascertain if a layout with third-party macros is printable as-is. To pass off sub-par designs as high quality, malicious third-party vendors may selectively modify their layouts to steer the detector to misclassify hotspot regions as non-hotspot. That is, attackers can hide hotspots in their low-quality macros by introducing adversarial perturbations.

Besides adversarial attacks, a stealthy poisoning attack is also threatening. It targets inserting backdoor in ML models during the training stage. Instead of requiring control over the model training process, this is achieved by poisoning the training data. A common poisoning mechanism is to insert a secret trigger to the features and coax ML models to unknowingly learn the secret trigger as a pattern of the attacker's target label. The work of~\cite{liu2020poisoning} demonstrates poisoning attacks on lithography problems.  

% Attackers may either directly poison samples with their target label or further change the labels of poisoned samples to their target ones.

\subsection{Attack Method on Model Performance}

Adversarial attacks are based on the generation of adversarial samples. The most fundamental attack method is fast gradient sign attack (FGSM)~\cite{goodfellow2014explaining}. For attacks without a specific target, it perturbs the input features $X$ towards the direction that maximizes the error $J$ between prediction $F(X|w)$ and the correct label $y$. This gradient ascent process is similar to the gradient descent operation on input in Equation~\ref{eq:input_opt}, but optimizes input $X$ towards the opposite direction. To avoid the attack being perceptual to victims, the perturbation is often constrained with a maximum perturbation amount $\epsilon$. For FGSM attack, the constraint $\epsilon$ is defined with an $l_{\infty}$ norm. The generation process of perturbed input $X_p$ is shown below. 

%When the constraint $\epsilon$ is defined with an $l_{\infty}$ norm, the FGSM attack process on generating perturbed input $X_p$ is shown below. 

\vspace{-.05in}
\begin{equation*}
    X_p \leftarrow \text{clip}(X + \epsilon \; \text{sign} (\nabla_{X} J(F(X|w), y)))
\end{equation*}
\vspace{-.05in}

Besides the fundamental FGSM, there are other adversarial attack methods like projected gradient descent (PGD)~\cite{madry2017towards}, which is a more effective, multi-step variant of FGSM.

These attack methods with FGSM or PGD are based on constraints limiting pixel-wise perturbation amplitude, viewing input as ordinary images. In ML for EDA, the scenario can be quite different, depending on the actual EDA application. When targeting lithography hotspot detectors~\cite{liu2020adversarial}, instead of perturbing every pixel, the perturbation in this case is to insert the shape of artificial sub-resolution assist features (SRAFs) to layouts. Potential attackers are low-quality IP/macros providers who wish to hide lithography deficiencies in their design or maliciously sabotage the downstream manufacturing process. 

In ML for EDA, adversarial attacks are more threatening to ML models targeting lithography problems, where design layouts as inputs can easily come from malicious third-party providers. In comparison, for models supposed to be deeply incorporated and coupled with existing design flows, like routability models, it will be more difficult for attackers to insert their perturbations to model inputs.

In addition, although we introduce adversarial attacks by assuming attackers have access to white-box model $F$ with weights $w$, actually they can also be applied to black-box scenarios. In this case, the adversarial samples can be generated based on certain surrogate models with similar functionality. These samples are still effective after transferring to the black-box target model $F$. This successful black-box attack attributes to the extraction of non-robust features by both surrogate model and target model $F$. Such non-robust features are features that are highly predictive, yet brittle and incomprehensible to humans~\cite{ilyas2019adversarial}. Allowing black-box scenarios further lowers the bar for adversarial attacks on ML model performance.

\begin{figure*}[!t]
%\twocolumn[{
  %  \vspace{-.15in}
\centering
    \subfigure[]{\includegraphics[height=0.184\textwidth]{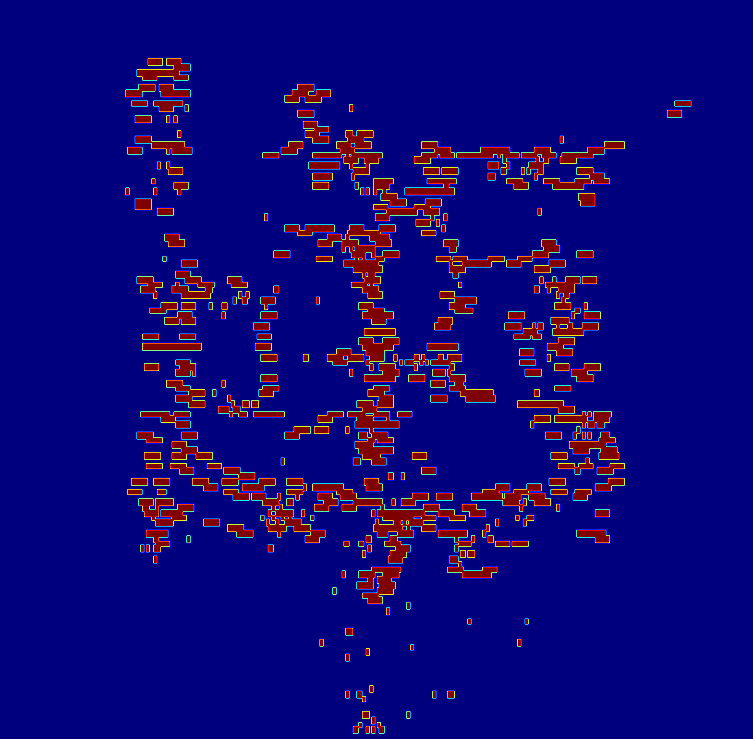}
    \label{attack1}}
    \subfigure[]{\includegraphics[height=0.184\textwidth]{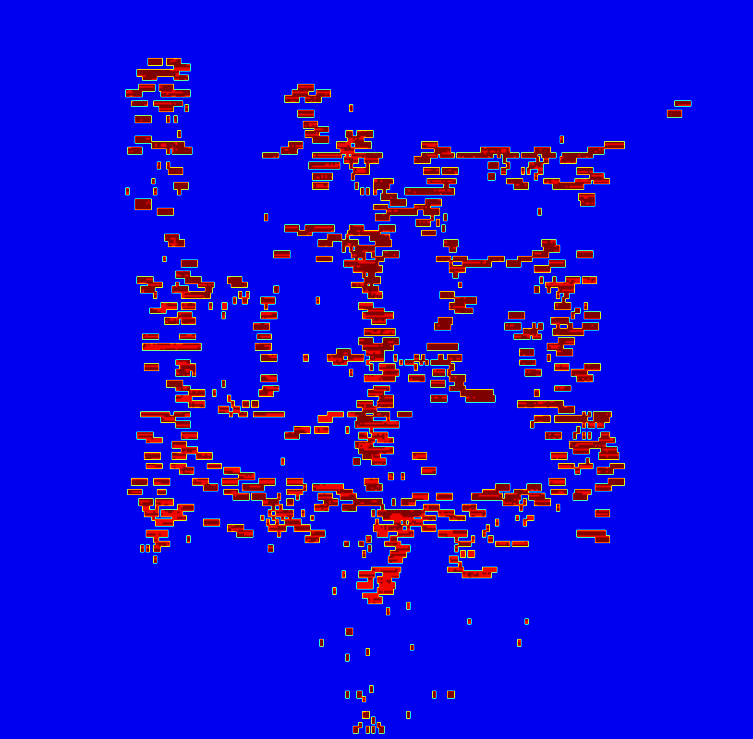}
    \label{attack2}}
    \subfigure[]{\includegraphics[height=0.184\textwidth]{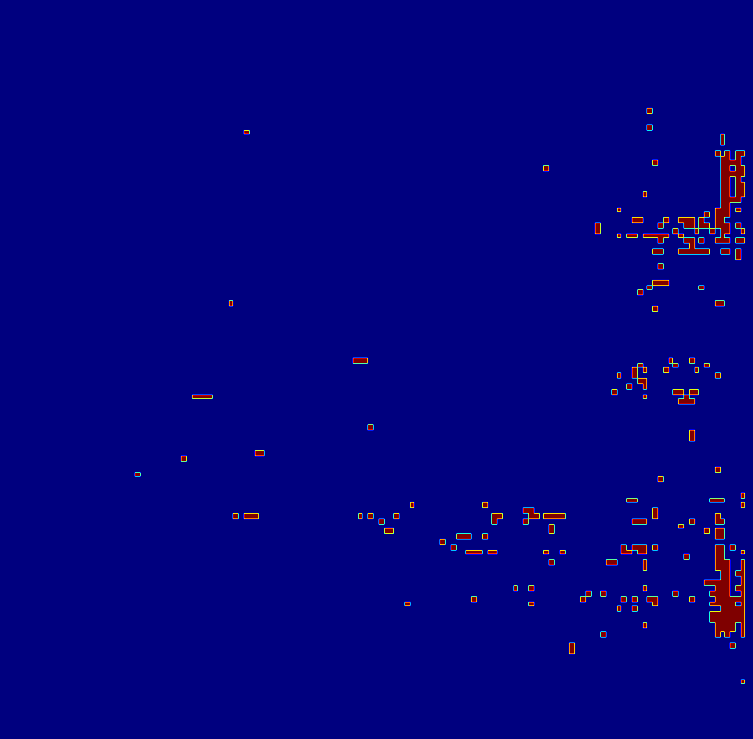}
    \label{attack3}}
    \subfigure[]{\includegraphics[height=0.184\textwidth]{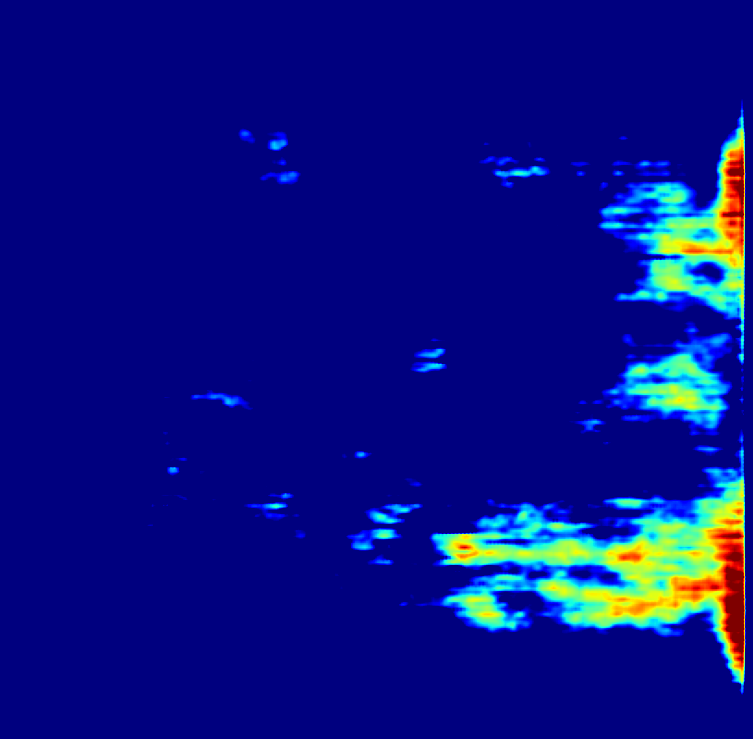}
    \label{attack4}}
    \subfigure[]{\includegraphics[height=0.184\textwidth]{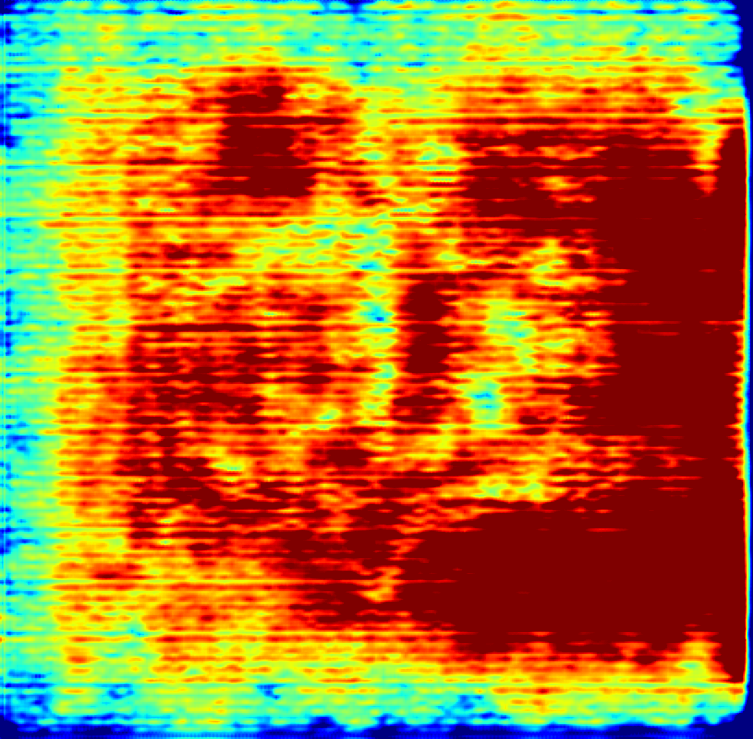}\label{attack5}}
    %\vspace{-.1in}
\caption{Adversarial attack example on routability prediction. (a) Original feature on the distribution of clock tree. (b) The clock tree feature after adversarial attack. The perturbation is small. (c) The congestion label. (d) The prediction based on original features in a. (e) The prediction based on adversarial attacked features in b. The attack makes predictions meaningless. }
    %\vspace{-.05in}
\label{adversarial_attack}
%    }]
\end{figure*}

Besides adversarial attacks, poisoning attacks target ML performance at the model training stage. Take the same lithography problem as an example, to hide lithography deficiencies, malicious insiders can stealthily introduce a backdoor into lithography detectors by providing poisoned training data with backdoor `triggers'. The detector is thus trained to link the trigger with non-hotspot. If this detector is adopted and deployed, any attackers knowing the backdoor can pass off a low-quality design as `hotspot-free' by inserting the trigger in their own layouts~\cite{liu2020poisoning}. Recent studies~\cite{liu2020bias} show that this poisoning attack on lithography can be defended by diluting the intentional bias from triggers with data augmentation strategies.

\subsection{Defense Method on Model Performance}
 
To cope with potential adversarial attacks in ML for EDA, we propose to build more robust models by adopting defense algorithms like curvature regularization (CURE)~\cite{moosavi2019robustness}. This work studies the relation between model \emph{curvature} and robustness against adversarial attacks. It first calculates the Hessian matrix on loss function with respect to input features, then proves that ML models with a smaller curvature (i.e. smaller eigenvalues of the Hessian matrix) demonstrate higher robustness~\cite{moosavi2019robustness}. Intuitively, smaller eigenvalues of Hessian indicate a smaller curvature around input, implying a `locally linear' behavior in the neighbor of input.

% is beneficial to obtain 

Therefore, to build more robust models, a solution is to penalize large eigenvalues of the aforementioned Hessian matrix with respect to the input. It is achieved by imposing gradient regularity (i.e., small curvature) along the direction of gradient descent. This new regularizer $R$ with respect to original input $X$ is shown below, and is added to the original loss function $L$ with a controllable weight $\alpha$. 

\vspace{-.05in}
\begin{align}
 & R = || \nabla_X L(X + hz) - \nabla_X L(X)||^2 \nonumber \\
 & Loss = L + \alpha R     
 \label{equ:cure}
\end{align}
\vspace{-.05in}

\noindent where the vector $z \propto \text{sign} (\nabla_X L(X))$, $h$ is a sufficiently small value controlling the step size. This new regularizer $R$ penalizes an approximation of the second-order derivative, which represents curvature with respect to input $X$.

In this paper, we studied adversarial attacks on different ML for EDA tasks. More importantly, we apply the CURE method to construct more robust models with very limited accuracy loss in tasks like lithography hotspot detection. It can better defend the adversarial attack~\cite{liu2020adversarial} on hotspot detectors. %proposed in previous works~\cite{liu2020adversarial}. 

% This method prevents potential adversarial attack by inducing a decrease in the model curvature. 
%indicates that a smaller curvature (i.e. small eigenvalues of the Hessian matrix with ). 

\subsection{Experiment on Model Performance Attack}

We first verify the effectiveness of the widely-adopted adversarial attack algorithm like PGD~\cite{madry2017towards} on routability models. These traditional adversarial attacks turn out to work well in ML for EDA tasks. Experimental results are shown in Figure~\ref{adversarial_attack}.  Figure~\ref{attack1} shows the original feature of the clock tree together with all flip-flops in the layout, and Figure~\ref{attack2} shows the corresponding feature with perturbations. Their difference in major patterns is not obvious. Figure~\ref{attack3} shows the congestion label. The normal prediction based on original features in Figure~\ref{attack4} is close to ground truth in Figure~\ref{attack3}. However, the prediction based on features with perturbations in Figure~\ref{attack5} is almost meaningless. The distinction between prediction results in Figure~\ref{attack4} and Figure~\ref{attack5} clearly indicates the effectiveness of the traditional attacks like PGD in ML for EDA tasks, as demonstrated on this routability problem. Similar results are also observed for the FGSM attack in our experiment.

However, the difference between Figure~\ref{attack1} and ~\ref{attack2} is still perceptual to humans, indicating inferior attack quality compared with attacks on general images. There are at least two reasons. First, the model performs binary classification on each grid instead of multi-class classification, leaving fewer inter-class decision boundaries. Second, the input feature is also close to binary, indicating the existence of the clock tree elements. The simple feature also makes perturbations more uniform and obvious. 

\begin{table}[!b]
  \centering
    %\vspace{-.05in}
     \resizebox{0.83\linewidth}{!}{ \renewcommand{\arraystretch}{1.2}  
  \begin{tabular}{| c | c | c |}
  \hline
\multirow{2}{*}{Model}   &  Accuracy  &    Attack     \\
                         &   (AUC)    &  (Success / Total)      \\
  \hline
  \hline
%  \multirow{2}{*}{Vanilla Model}   & \multirow{2}{*}{0.895}  &   \multirow{2}{*}{0.171}   \\  
%        &     &  \\
    Vanilla Model  &  0.895   &   0.171 \\
  \hline
%  \multirow{2}{*}{Robust Model}   &  \multirow{2}{*}{0.885}  &  \multirow{2}{*}{0.136}   \\
%        &     &  \\
    Robust Model   &  0.885   &  0.146 \\
  \hline
  \end{tabular}
    }
    \vspace{.1in}
  \caption{Adversarial attack and defense on lithography hotspot detectors. The attack is performed by inserting SRAFs on inputs. The robust model based on CURE regularizer reduces the attack success rate with limited accuracy loss. }
  \label{tbl:adversarial}
   %   \vspace{-.15in}
\end{table}

As mentioned, adversarial attacks in ML for EDA are more threatening in a few special tasks, like for lithography problems. The corresponding adversarial attack constraint is also different. To study this task, we first replicate the adversarial attack in the work of~\cite{liu2020adversarial} on the same dataset. It attacks lithography hotspot detectors by inserting artificial SRAFs as perturbations. The accuracy of this model and attack success rate on it are shown in the `vanilla model' of Table~\ref{tbl:adversarial}. After that, we apply the CURE regularizer in Equation~\ref{equ:cure} to construct a more robust model. As the comparison in Table~\ref{tbl:adversarial} shows, the attack success rate drops from 0.171 to 0.146 on our robust model, while the accuracy slightly degrades from 0.895 to 0.885. It indicates the CURE-based robust model is less vulnerable to adversarial attacks in this specific task with very limited accuracy loss. In the future, we will further explore more robust models by customizing the CURE method to the constraint in SRAF shapes for this task.

%while the accuracy loss for ordinary inputs is very limited.  

\section{Unreliability in ML Performance}
\label{sec:unreliability}

\subsection{Model Unreliability Overview}

%The first type of security concern is on ML models themselves. The security problems of ML models in EDA can be categorized into two major types, 1) unauthorized access to ML models; 2) reliability problems of ML models. 
%Both of them can cause serious accuracy loss in ML-integrated circuit design flow. 

We have discussed three major types of security concerns in ML for EDA, targeting data privacy, ML model competitive advantage, ML model performance, respectively. They are all malicious attacks.  
Since we have defined the `security' to include all measures related to unforeseen consequences, we emphasize one additional security concern, which is model unreliability. It is reflected by observations that model accuracy may seriously degrade on some testing samples. It is not caused by any malicious attackers, but can be especially serious in ML for EDA, because of several special properties.

% of ML for EDA. 
%We discuss them in detail below. 

\begin{figure}[!b]
\centering
    %\vspace{-.15in}
    \includegraphics[width=.99\linewidth]{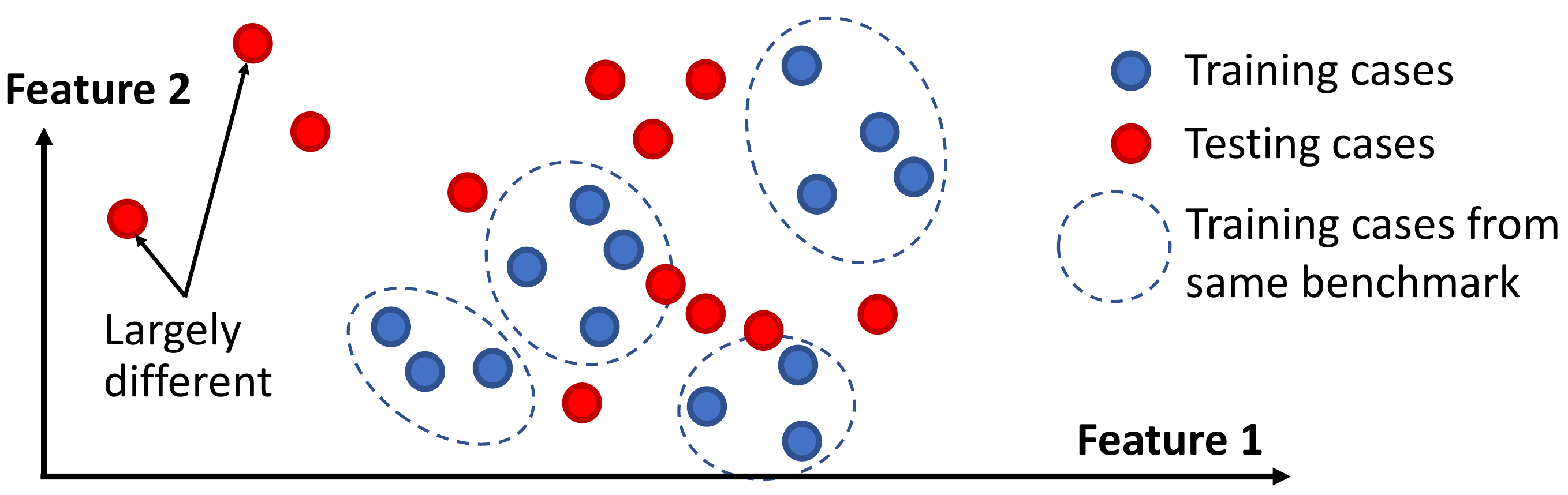}
  %  \vspace{-.1in}
\caption{Distribution of training and testing samples.}
\label{distribution}
%\vspace{-.15in}
\end{figure}

First, as mentioned in Subsection~\ref{subsec:property}, huge heterogeneity may exist between training and testing data samples, resulting from the large difference among circuits due to functionality, micro-architecture, and technology node. 
We further illustrate this concept in Figure~\ref{distribution}, which shows a possible distribution of training and testing samples in a simplified input feature space. A few testing samples can be largely different from existing training samples, inevitably leading to accuracy degradation in these samples. In this example with only two features, the difference between samples seems obvious and easy to measure. But in practice, such `difference' and the corresponding impact on testing accuracy are very difficult to know. 

%As mentioned, models trained on designs like Cortex-M0 with 90nm technology may not perform well on latest processors, like Neoverse N2 with 5nm technology. This level of training data heterogeneity is uncommon in benchmarks for general ML applications like computer vision. 

Second, it is very difficult for engineers to be aware of unexpected accuracy degradation on a few testing samples in practice. Accurate detection of accuracy degradation requires ground-truth labels to be collected, which is highly time-consuming and inherently against the purpose of adopting ML models. Undetected accuracy degradation can lead to much less optimized design solution or even chip failure, causing serious income loss for users from design companies.

%Third, the high complexity in input data and pattern exacerbates the previous two properties. A circuit consists of orders-of-magnitude more information than ordinary images. For prediction tasks, models are learning behaviors of complex EDA engines. For optimization tasks like macro placement, models are exploring a huge solution space~\cite{mirhoseini2021graph}, much larger than a Go game. 

Third, as mentioned in Subsection~\ref{subsec:property}, high complexity exists in both input data and the pattern for ML models to learn. These complexities exacerbate the difficulty in the study of input sample similarity or the detection of possible accuracy degradation.

In practice, the unreliability problem can be reflected by concerns like: `Does the ML model work on 7nm technology or memory/GPU/certain IPs? To what extent may the accuracy degrade? Is transfer learning on new data necessary?' Currently this is mostly speculated based on model developers' confidence and intuition. To the best of our knowledge, there is no systematic study on this topic. As a result, users cannot safely trust any ML model in EDA before they have a deep understanding of the potential unreliability in model performance. It affects all four scenarios we mentioned in Table~\ref{tbl:scenario} and may become a major obstacle that prevents a wide application of ML in EDA in the future.

% from users. For example, we often receive questions

%\subsection{Reliability of ML Models}
%There are many possible sources of unreliability in ML models. They can be mainly caused by the design data heterogeneity, from the large difference among circuits due to micro-architecture, functionality, and technology node. Another possible degradation of ML models. It is possible to improve the reliability of ML models by .

%\subsection{Defense Algorithms or Guidelines}

\subsection{Model Unreliability Analysis}

For each ML model in EDA, understanding `unreliability' requires detecting accuracy loss without knowing the label, or quantitatively determining the appropriate scope of testing samples. This solution is not straightforward. One direction we can think of is to define a new metric to measure the similarity between training data and each testing sample. As Figure~\ref{distribution} indicates, the performance unreliability (degradation) is mostly caused by the sparse distribution of data samples in the feature space. If the similarity between one testing sample and the model training data is lower than a certain threshold, the ML model should either reject inference on this testing sample or at least raise a warning. Another direction is to adopt ML models with prediction confidence incorporated in their prediction outputs. Low confidence generally indicates uncertainty and possible accuracy degradation on the testing sample. The confidence is available as probability values for many classifiers, especially multi-class classifiers, but less obvious in common regression tasks in ML for EDA.

%For confidence lower than threshold, . 

%If we can detect testing accuracy or define the appropriate scope of testing samples, given a dataset, it is possible for us to construct multiple ML models. 

Understanding `unreliability' in ML models not only helps to avoid unexpected accuracy drop, but also provides guidance during model construction. If we can detect testing accuracy or define the appropriate scope of testing samples, then given a dataset, it is possible for developers to construct multiple ML models, each trained with part of training data and applied to a specific scope of testing samples. For example, in Figure~\ref{distribution}, developers may train one ML model based on each benchmark, instead of training one general model with all samples in the training dataset. By applying different models to different testing samples, better overall results can be achieved.

%than training a general model with all samples in the training dataset.  
%another related question is, given existing . 
%the testing sample can be .  

\begin{figure}[!t]
\centering
    \includegraphics[width=\linewidth]{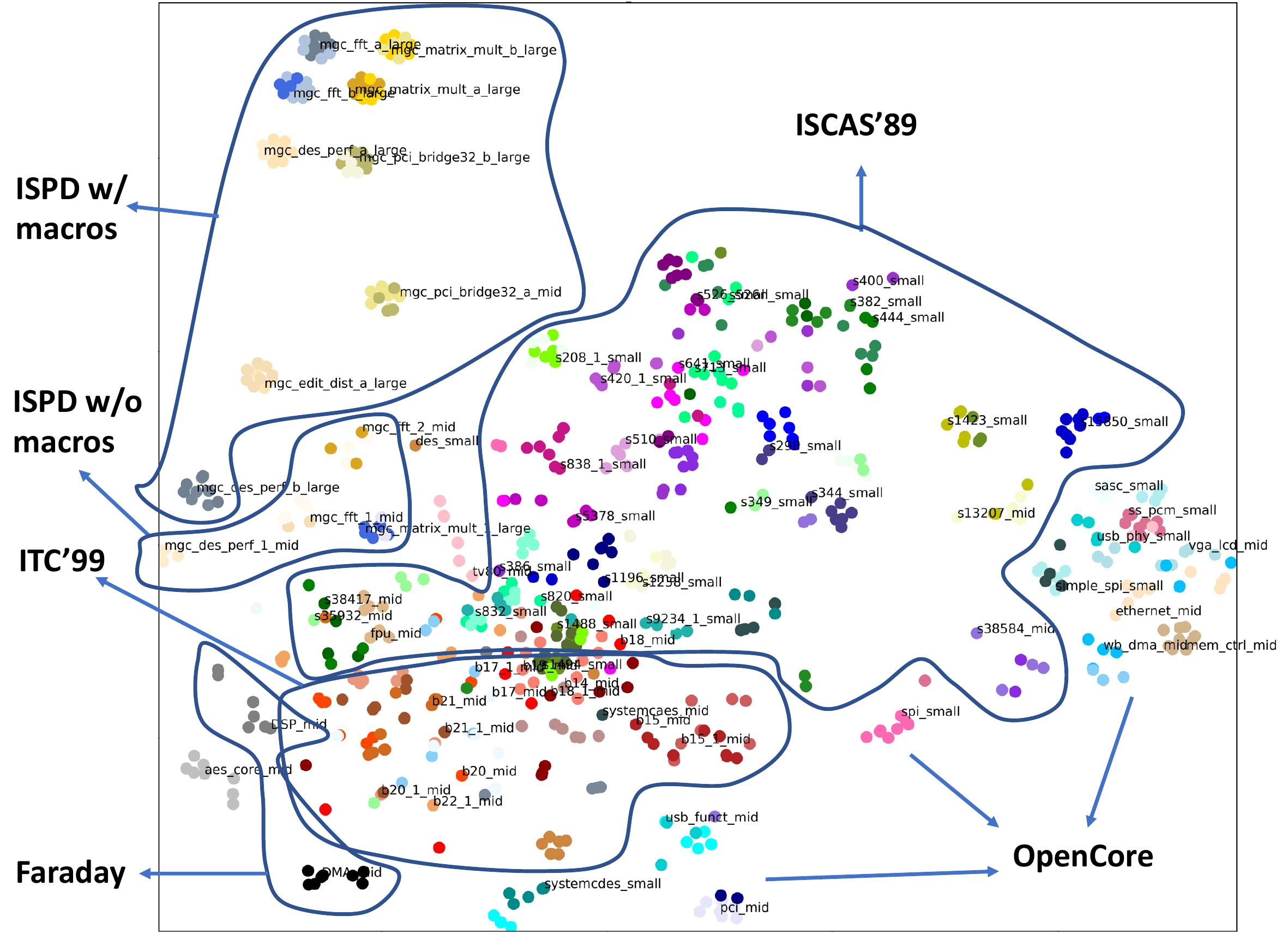}
    \vspace{.05in}
\caption{Visualization of designs/layouts by dimension reduction. Each point represents one layout and same color indicates the same design. Similarities can be observed for designs/layouts from the same benchmark.}
\label{clustering}
\vspace{-.05in}
\end{figure}

\subsection{Experiment on Unreliability and Data Similarity}

We first demonstrate the accuracy degradation when applying models on largely different designs, and present our preliminary study in understanding the design similarity and model performance.

Accuracy degradation on specific testing samples is very common during the development of ML models. For example, while a carefully designed routability model on congestion prediction achieves an average performance of 0.83, its performance can be lower than 0.70 for a few testing designs.

To demonstrate the idea of measuring `similarities' between design and samples, we try to visualize multiple layouts from various designs in different benchmarks in the routability experiment. We adopt simple principle component analysis (PCA)~\cite{wold1987principal}-based dimension reduction techniques. The visualization is shown in Figure~\ref{clustering}. Each point in this figure indicates one layout solution and same color indicate layouts from the same design. Different benchmark names are annotated on the figure. To provide more information, the tiny text annotated on some points is in the format of design name plus the size ('small', 'mid', 'large') of design. For example, the tiny text `pci\_mid' at bottom purple points of Figure~\ref{clustering} indicates design name `pci' with middle-level layout size.

We can observe very interesting and reasonable clustering of layouts and designs in Figure~\ref{clustering}. First, layouts from the same design with the same color are very closely clustered. Second, intuitively similar designs, like designs from the same benchmark, are obviously closer to each other. For example, all designs from the ISPD benchmark distribute on the upper left corner of Figure~\ref{clustering}. More importantly, the designs with macros are clearly closer to the corner, showing a larger difference with most designs without macros. Similarly, designs from ITC'99 and small designs from ISCAS'89 reflect clear intra-benchmark similarities. 
Studies like this can provide guidance in quantitative measurements of design similarity and understanding of model unreliability. A straightforward example is, models trained with small design layouts (in the center of Figure~\ref{clustering}) may not perform well on large designs with macros (in the upper left corner).

As mentioned, such design similarity also provides guidance in model construction. We provide an experiment on developing DRV detection models with different training data in Table~\ref{tbl:multi_model}. Notice that this preliminary experiment targets DRV, thus overall accuracy tends to be lower than congestion detection in previous experiments. In this experiment, all training and testing designs are classified into three types: small, middle, large, according to their layout size. Then we try to train models either on all training data or on part of training data. As Table~\ref{tbl:multi_model} shows, the model trained on all designs does not perform the best. Instead, the model trained only on middle designs performs better on middle and large testing designs. This preliminary result supports our speculation that based on design similarity, constructing multiple ML models for different testing scopes can achieve better accuracy.

\begin{table}[!t]
  \centering
    \vspace{.1in}
\resizebox{0.94\linewidth}{!}{     \renewcommand{\arraystretch}{1.1}  
  \begin{tabular}{| c || c | c | c | }
  \hline
   \multirow{3}{*}{Test on}  &  \multicolumn{3}{c|}{Training on}  \\
   \cline{2-4}
                & \multirow{2}{*}{Middle}  &  \multirow{2}{*}{Small + Middle}  &  Small + Middle    \\  
                             &          &                  &  + Large (All) \\
  \hline
  \hline
  Small        &  70.6  &  \textbf{72.4}  &  71.5  \\
  \hline
  Middle       &  \textbf{75.6}  &  75.4  &  71.3  \\
  \hline
  Large        &  \textbf{71.3}  &  64.9  &  71.0  \\
  \hline
  \end{tabular}
  }
  \vspace{.05in}
 \caption{Testing accuracy of DRV detection models. Three models are trained with different partitions of data.}
   \vspace{-.05in}
  \label{tbl:multi_model}
\end{table}
% zx52@hl279-cmp-02:~/OOD/week2/method_ml

%detect such unexpected inaccuracies without knowing the label. 

\iffalse
\section{Summary of All Security Concerns}
\label{sec:summary}

\begin{table}[!h]
  \centering
  \caption{Summary of Security Concerns}
  \label{tbl:overview_concern}
  \vspace{-.1in}
  \begin{tabular}{l | c c c}
 	\hline
    Security Concerns  &  Practicality  & Uniqueness \\        
	\hline
	1. Data Privacy              &    Low       &    Low     \\
	2. Model Competitiveness     &    High      &    Low  \\
	3. Model Performance         &    Medium    &    High    \\
	4. Model Unreliability       &    High      &    High    \\
	\hline
  \end{tabular}
      \vspace{-.05in}
\end{table}

Table~\ref{tbl:overview_concern} gives a summary of all major security concerns covered in this paper. We qualitatively inspect these concerns based on two criteria named `practicality' and `uniqueness'. First, we are interested in whether each concern is practical to apply in realistic. Second, we inspect whether the concern is unique in ML for EDA, or similar to existing security studies in general ML tasks.

First, the malicious attack on training data privacy, as Section~\ref{sec:1} indicates, turns out to be quite difficult and relies on many strong assumptions. Also, the attack methodology turns out to be similar to most model inversion attacks in general ML tasks. Second, as Section~\ref{sec:2} indicates, the attack on model competitive advantage is easy to implement. Our current method is also similar to most model extraction methods.  

\fi

\section{Potential Concerns in the Future}
\label{sec:end}

We have presented four major types of security concerns in ML for EDA. At the end of this study, we try to further anticipate a few other potential concerns or impacts that may arise in the future, when ML for EDA becomes more ubiquitous.

%. They are not problematic now, but may arise in the future, when ML for EDA becomes ubiquitous.  
%As this new technology keeps emerging in this field, 

\subsection{Security in Decentralized Setting}

\begin{figure}[!t]
    \centering
    \includegraphics[width=0.9\linewidth]{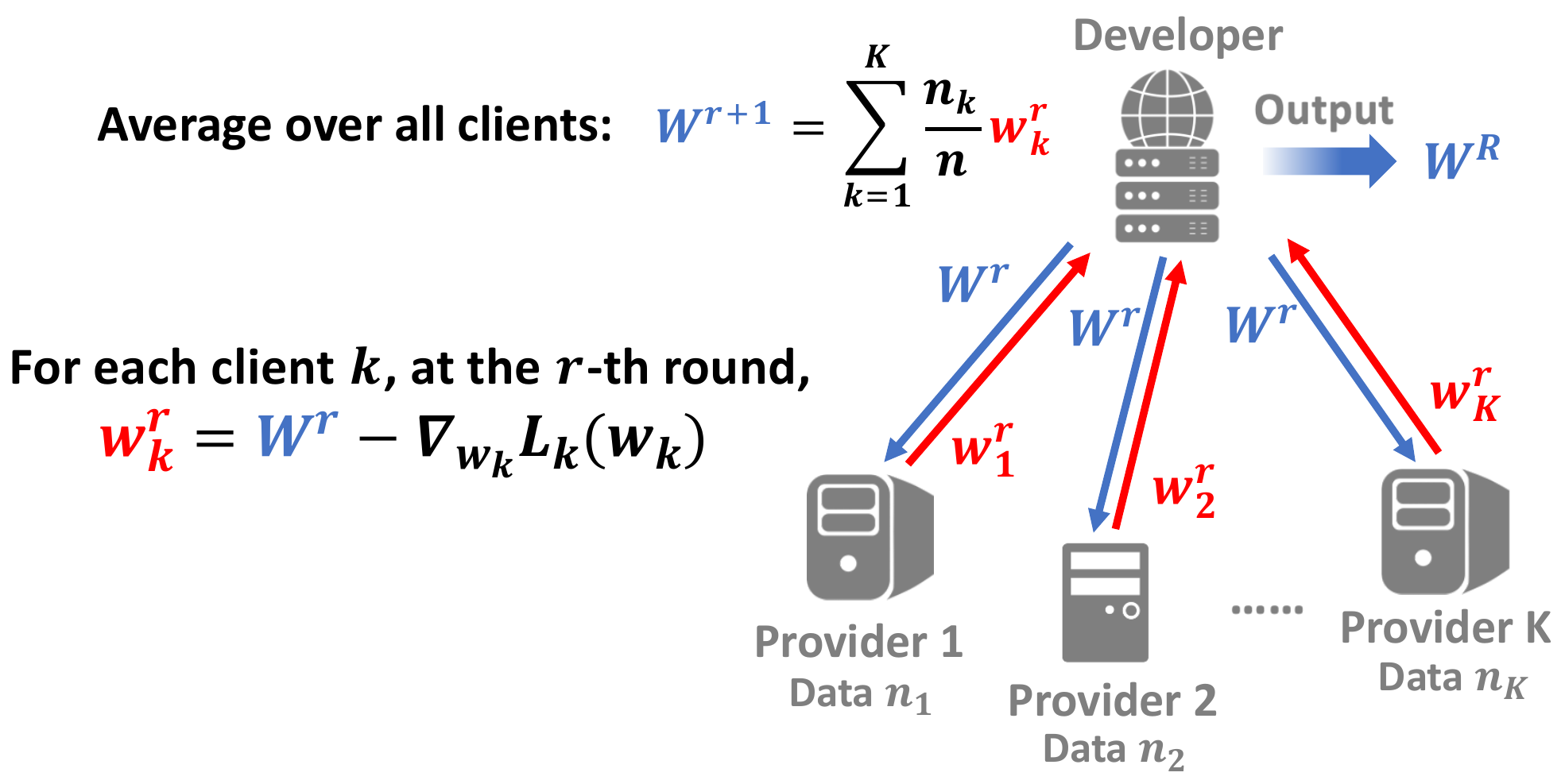}
    \vspace{-.05in}
    \caption{The visualization of the decentralized training.}
    \label{fig:decentralized}
    \vspace{-.15in}
\end{figure}

The effectiveness of ML for EDA largely hinges on the availability of a large amount of high-quality training data. In reality, developers have very limited access to the latest design data, which is owned by design companies and mostly confidential. 
Such data availability problem is becoming the limiting constraint on the future growth of ML for EDA and chip design. Considering the decentralized distribution of high-quality circuit data, we have witnessed explorations~\cite{pan2022towards} based on federated learning (FL), as Figure~\ref{fig:decentralized} illustrates. Developers collaboratively train one ML model based on the private local data from $K$ data providers. In each round, data providers send locally trained models to the central server, then the server aggregates and distributes the updated model back to all providers. This may become a major trend in constructing and deploying ML models in the future. 

%Although one can commission ML model training to a design company, the data of a single company might be still inadequate or biased, especially for small companies. 

However, collaboratively constructing ML models in a decentralized setting incurs many new security concerns. For example, if one of the data providers is an attacker, it leads to serious security threats. First, the attacker can directly get full access to the trained ML model during the collaborative training process. Second, the attacker can easily insert malicious backdoor attacks into the ML model, by including poisoning samples in its local training dataset. Third, it is possible for the attacker to recover the private data of other data providers. 
This can be achieved based on the idea of generative adversarial networks (GAN). The attacker can use the trained model as a discriminator, and train an additional generator to recover input samples of a specific class~\cite{hitaj2017deep}. But this requires the model to be a multi-class classifier, which is not common in ML for EDA tasks.  

% (recover) 
%Third, the attacker can get unauthorized access to other clients. 

\subsection{Label Generation with ML Models}

As mentioned in Subsection~\ref{subsec:competitiveness}, label generation is one of the most costly steps in model development. In the future when ML models for EDA achieve higher accuracy and become ubiquitous, it is possible to directly apply existing models to generate training labels for the development of new ML models. While this greatly reduces label generation cost, accuracy degradation is unavoidable.   

The accuracy degradation can already be observed in the model extraction experiment in Subsection~\ref{subsec:competitiveness_exp}. As Table~\ref{tbl:model_compete} shows, the attack model trained on 40\% unlabeled data is less accurate than the original victim model trained on 40\% labeled data. Model developers should be aware of such accuracy loss and avoid overuse of pseudo-labels.

\subsection{Impact on EDA Tools}

In the future, models learning the functionality of EDA tools may be applied to partially or even entirely replace these EDA tools in circuit design flow. This is a quite special impact. Different from most ML applications where models replace human efforts, ML for EDA methods have been applied to accomplish the tasks of both human designers and EDA tools.

To avoid emerging competition with their own tools, in the future EDA vendors may hope to revise existing user license agreements and prevent unauthorized use of their tools to develop ML models with similar functionalities. However, disabling model training based on a specific software is technically very difficult, and violations of this rule cannot be easily detected by the vendor.

%To the best of our knowledge, this is not explicitly studied before. Here we contribute a few directions that worth exploration.

\section{Conclusion}

In this paper, we provide a comprehensive and impartial summary of all safety concerns we observe in ML for EDA tasks. According to our study, some concerns like model extraction, attacks on model performance, and inherent model unreliability are highly threatening, while potential design privacy attack turns out to be less practical. In the future, we will explore more customized attack and defense methodologies with more in-depth experiments in ML for EDA. 

%more in-depth studies will be conducted. 

%This paper provides our preliminary analysis results based on straightforward .  

%Most of these concerns are neglected by practitioners and researchers.   

%=======================================%
%=======================================%

%=======================================%
%               Bibliography            %
%=======================================%

\bibliographystyle{IEEEtran}
\bibliography{main.bib}
%=======================================%
%=======================================%
\end{document}